# Josephson systems based on ballistic point contacts between single-band and multi-band superconductors

Y.S. Yerin[1,2], A.S. Kiyko[1], A.N. Omelyanchouk[1], E. Il'ichev[3]

[1]*B.Verkin Institute for Low Temperature Physics and Engineering of the National Academy of Sciences of Ukraine 47 Lenin Ave., 61103 Kharkov, Ukraine*

[2]*Institute for Physics of Microstructures RAS, GSP-105, Nizhny Novgorod, 603950, Russia*

[3]*Leibniz Institute of Photonic Technology, D-07702 Jena, Germany*

## Abstract

The Josephson effect in ballistic point contacts between single-band and multi-band superconductors was investigated. It was found that in the case of Josephson junctions formed by a single-band and an $s_\pm$-wave two-band superconductor as well as by a single-band and a three-band superconductor the junctions become frustrated, demonstrating the φ-contact properties. Depending on the ground state of a three-band superconductor with broken time-reversal symmetry (BTRS), the Josephson junction can have from one to three energy minima, some of which can be locally stable. We also study the behavior of a dc SQUID based on the Josephson junctions between single-band and multi-band superconductors. Some features on the dependences of the critical current and the total magnetic flux on the applied flux of a dc SQUID based on the Josephson point contacts between a single-band superconductor and an $s_\pm$-wave superconductor, three-band superconductor with BTRS and three-band superconductor without BTRS as compared to the conventional dc SQUIDs based on single-band superconductors were found. The results can be used as an experimental tool to detect the existence of multi-band structure and BTRS.

## I. Introduction

One of the most efficient ways of obtaining information about the symmetry of the order parameter in unconventional superconductors is phase-sensitive techniques based on the Josephson effect in such superconducting systems. Armed with the hypothesis about a possible form of the Cooper pair wave function, we can theoretically predict specific aspects of the behavior of various Josephson systems based on unusual superconductors. These features form the basis for the technique known as Josephson interferometry. This technique involves the study of the magnetic response of a Josephson junction, current-phase relation for the Josephson junctions formed at grain boundaries and SQUID interferometry. Among the variety of the Josephson interferometry techniques, the latter is often the most useful. From a technical point of view, it is based on the study of the characteristics of a one-contact interferometer (a Josephson junction in a superconducting ring) or a dc SQUID (a SQUID ring with two Josephson junctions). In this geometry, one of the superconductors forming the junction has an isotropic s-wave symmetry of the order parameter, while the second one is an unusual superconductor with the symmetry of the order parameter to be revealed [1,2].

It is important to note that the Josephson interferometry has already proved itself as a useful technique that greatly helped in the identification of *d*-wave pairing mechanism of Cooper pairs in cuprate high-$T_c$ superconductors (see, e.g., review Ref. 3).

The recent discovery of a new class of high-$T_c$ iron-based superconductors [4] gave rise to the question of the pairing mechanism in these compounds and hence the symmetry of the superconducting order parameter. The initially widely accepted hypothesis of the so-called sign-reversal two-component $s_\pm$-wave order parameter [5,6] does not allow us to unambiguously explain the experimental data for some of the iron-based superconductors [7–14]. In this regard, there appeared models based on the assumption of a multi-component chiral structure of the order parameter with the symmetry of the with the symmetry of types $s+id$ [15], $s_\pm+is_{++}$ [16], or with conventional *s*-wave symmetry with three or more gaps. Under certain conditions, the presence of chirality leads to the appearance of frustration, when the time-reversal symmetry in a superconductor is broken [17–26]. This means that the phases of the order parameter cannot simultaneously satisfy the minimum energy condition, hence creating a two or more fold degenerate ground state.

According to theoretical predictions, broken time-reversal symmetry in iron superconductors should lead to some interesting phenomena, such as the appearance of spontaneous magnetic field in the presence of nonmagnetic defects, massless Leggett modes and phase solitons (see review Ref. 27 and references therein).

While several technologies have already been proposed for the detection of potential BTRS in iron-based superconductors [28–31] currently there is no strong experimental evidence for the existence of this phenomenon in iron-based superconductors.

Since the BTRS is, as a matter of fact, a consequence of the frustration of the order parameter phases, it is logical to assume that for the detection of this phenomenon, the above mentioned phase-sensitive experiments, and in particular the Josephson interferometry, can be advantageous.

Previously, we investigated the Josephson effect [32] and the behavior of a dc SQUID with Josephson point contacts between an s-wave superconductor and a three-band isotropic superconductor in the dirty limit [30] We have revealed the unusual dependence of the critical current on the external magnetic flux and demonstrated the possibility of appearance of multi-hysteresis loops in the dependence of the total flux on the external magnetic flux. It was found that all these features of Josephson systems are associated with BTRS. Therefore, the Josephson interferometry is a powerful tool for the detection of this phenomenon. To complete the picture of the possibilities of using the Josephson interferometry to determine the structure of the order parameter and the possible BTRS in iron-based superconductors, in the present paper we will consider the Josephson effect and the behavior of a dc SQUID in another limiting case, when the Josephson point contact between a single-band and a multi-band superconductor has ballistic type of conductivity.

## II.     Formalism

The theory of stationary Josephson effect in ballistic point contacts (S-c-S contacts) formed by s-wave single-band superconductors has been developed in Ref.33. The results of this study can be generalized to the point contact between a single-band and an *n*-band ($n \geq 2$) superconductor (Fig. 1).

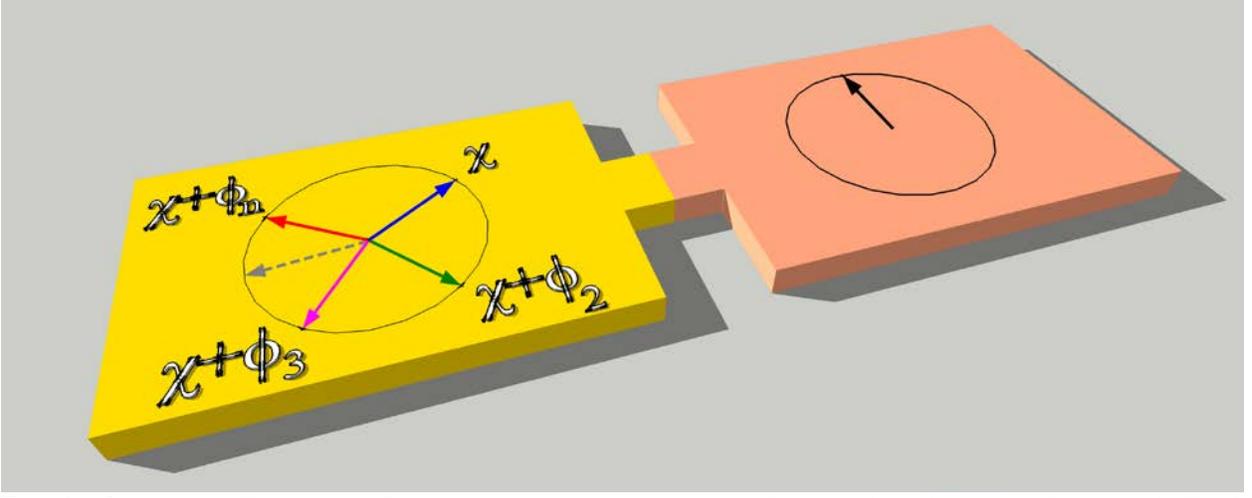

Fig. 1. Schematic illustration of a point contact between single-band (coral) and multi-band (yellow) superconductors. The length of the point contact much greater than its width, and the width is much smaller than the minimum value of the coherence length and the London penetration depth for a single-band superconductor and for the $i$-th band of a multiband superconductor.

In this case the total current through the Josephson contact at any temperature $T$ is given by the expression:

$$I = \sum_{i=1}^{n} \frac{2\pi |\Delta_0||\Delta_i|}{eR_{Ni}} \sin\left(\chi + \phi_i \operatorname{sgn}(i-1)\right) T \\ \times \sum_{\omega>0} \left( \frac{1}{4}\left[\left(\sqrt{\omega^2+|\Delta_0|^2} + \sqrt{\omega^2+|\Delta_i|^2}\right)^2 - (|\Delta_0|+|\Delta_i|)^2\right] + |\Delta_0||\Delta_i|\cos^2\frac{\chi + \phi_i \operatorname{sgn}(i-1)}{2}\right)^{-1} \quad (1)$$

Here $\chi$ is the phase difference between the first order parameter of the multi-band superconductor and the single-band superconductor, $\phi_i = \varphi_i - \varphi_1$ denotes the phase difference between $i$-th and the first order parameter of the bulk $n$-band superconductor, $|\Delta_0|$ is the energy gap of the single-band superconductor, $|\Delta_i|$ are values of the energy gaps of the $n$-band superconductor, $\omega$ is Matsubara frequency and $R_{Ni}$ are partial contributions of each band to the total resistance of the contact in the normal state.

The temperature dependence of energy gaps in the $n$-band superconductor can be found by the numerical solution of the self-consistency equation

$$|\Delta_i| = 2\pi T \sum_i \sum_j \sum_{\omega>0} \lambda_{ij} \frac{|\Delta_i|\exp\left[\mathrm{I}\phi_i \operatorname{sgn}(i-1)\right]}{\sqrt{\omega^2+|\Delta_i|^2}}, \quad (2)$$

where $\lambda_{ij}$ are the constants of electron interactions of the $n$-band superconductor, I is the imaginary unit.

Let us make few comments regarding the phase differences $\phi_i$, which determine ground states of a $n$-band superconductor. For a bulk clean two-band ($n = 2$) superconductor the ground state is non-degenerate with $\phi_2 = \phi = 0$ or $\phi_2 = \phi = \pi$, respectively, depending on the character of the interband interaction (attraction or repulsion). As it was shown in Refs. 31 and 32 in the case of a three-band superconductor degree of the degeneracy of the ground state is determined by the values of the interband interaction coefficients and modules of order parameters. Within the microscopic description it was found that the phase differences of the order parameters $\phi_2 = \phi$ and $\phi_3 = \theta$ are if $\phi \in \left[-\dfrac{\pi}{2}, \dfrac{\pi}{2}\right]$ and $\theta \in \left[-\dfrac{\pi}{2}, \dfrac{\pi}{2}\right]$, then

$$\begin{cases} \phi = \pm \arcsin \Omega, \\ \theta = \mp \arcsin\left(\dfrac{G_1 |\Delta_2|}{G_3 |\Delta_3|}\Omega\right), \end{cases} \quad \begin{cases} \phi = 0, \\ \theta = 0, \end{cases} \tag{3}$$

if $\phi \in \left[\dfrac{\pi}{2}, \dfrac{3\pi}{2}\right]$ and $\theta \in \left[-\dfrac{\pi}{2}, \dfrac{\pi}{2}\right]$, then

$$\begin{cases} \phi = \pi \pm \arcsin \Omega, \\ \theta = \pm \arcsin\left(\dfrac{G_1 |\Delta_2|}{G_3 |\Delta_3|}\Omega\right), \end{cases} \quad \begin{cases} \phi = \pi, \\ \theta = 0, \end{cases} \tag{4}$$

if $\phi \in \left[-\dfrac{\pi}{2}, \dfrac{\pi}{2}\right]$ and $\theta \in \left[\dfrac{\pi}{2}, \dfrac{3\pi}{2}\right]$, then

$$\begin{cases} \phi = \pm \arcsin \Omega, \\ \theta = \pi \pm \arcsin\left(\dfrac{G_1 |\Delta_2|}{G_3 |\Delta_3|}\Omega\right), \end{cases} \quad \begin{cases} \phi = 0, \\ \theta = \pi, \end{cases} \tag{5}$$

if $\phi \in \left[\dfrac{\pi}{2}, \dfrac{3\pi}{2}\right]$ and $\theta \in \left[\dfrac{\pi}{2}, \dfrac{3\pi}{2}\right]$, then

$$\begin{cases} \phi = \pi \pm \arcsin \Omega, \\ \theta = \pi \mp \arcsin\left(\dfrac{G_1 |\Delta_2|}{G_3 |\Delta_3|}\Omega\right), \end{cases} \quad \begin{cases} \phi = \pi, \\ \theta = \pi. \end{cases} \tag{6}$$

Here $\Omega = \sqrt{1 - \left(\dfrac{G_3^2 G_2^2 |\Delta_3|^2 - G_1^2 G_3^2 |\Delta_1|^2 - G_1^2 G_2^2 |\Delta_2|^2}{2 G_1^2 G_2 G_3 |\Delta_1||\Delta_2|}\right)^2}$, $G_1 = \lambda_{12}^{-1} N_1 + \lambda_{21}^{-1} N_2$, $G_2 = \lambda_{23}^{-1} N_2 + \lambda_{32}^{-1} N_3$

and $G_3 = \lambda_{13}^{-1} N_1 + \lambda_{31}^{-1} N_3$, $N_i$ are density of states at the Fermi level for each of the bands. The selection of stable solutions corresponding to the frustrated or non-frustrated state is determined by the second variation of the energy of the three-band superconductor with the difference of the phases $\phi$ and $\theta$.

In order to simplify the analysis of the problem we make a few assumptions. Let the temperature $T$ be equal to zero and assume that superconducting energy gaps $|\Delta_0|=|\Delta_i|=|\Delta|$ are equal. These assumptions will help us to understand qualitatively the main features of the behavior of the Josephson system without the complex numerical solution of Eqs. (2) - (6) in the general case.

Based on these assumptions, from Eq. (1) we have an expression for the current flowing through the Josephson contact

$$I = \sum_i \frac{\pi |\Delta|}{eR_{Ni}} \sin \frac{\chi + \phi_i \operatorname{sgn}(i-1)}{2} \operatorname{sgn} \cos \frac{\chi + \phi_i \operatorname{sgn}(i-1)}{2}, \tag{7}$$

And after integration over $\chi$ we find the energy of the Josephson junction

$$E = -\sum_i \frac{\Phi_0 |\Delta|}{2eR_{Ni}} \left| \cos \frac{\chi + \phi_i \operatorname{sgn}(i-1)}{2} \right|. \tag{8}$$

### III. Josephson effect in point contacts between single- and multi-band superconductors

As follows from Eqs. (7) and (8) the total current flowing through the Josephson junction between tw $s$-wave single-band superconductors, is [30]

$$I = \frac{\pi |\Delta|}{eR_{Ni}} \sin \frac{\chi}{2} \operatorname{sgn} \cos \frac{\chi}{2}, \tag{9}$$

and the energy of Josephson junction

$$E = -\frac{\Phi_0 |\Delta|}{2eR_{N1}} \left| \cos \frac{\chi}{2} \right|. \tag{10}$$

The current-phase relation of this contact and the dependence of the Josephson energy on the phase difference $\chi$ is shown in Fig. 2.

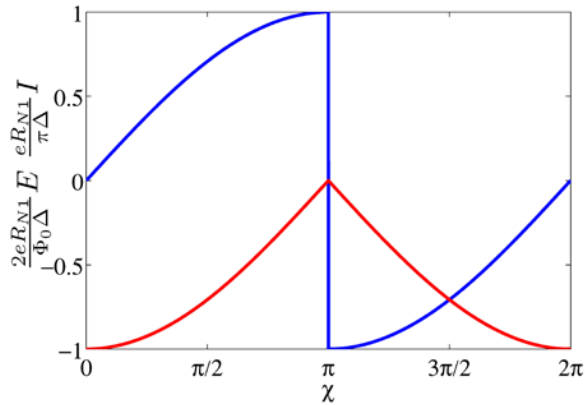

Fig. 2. The current-phase relation (blue line) and the Josephson energy (red line) of a point contact between two single-band superconductors.

The dependence $I(\chi)$ undergoes a jump at the point $\chi = \pi$. This jump is the main difference between the Josephson junction with ballistic conductivity and the similar system with diffusive conductivity [34].

In case of a junction formed by single-band and two-band superconductors the current-phase relation can acquire new qualitative features, if the two-band superconductor has $s_\pm$-wave symmetry of the order parameter (the characteristics of the Josephson junction with a two-band superconductor of $s_{++}$-wave symmetry are qualitatively similar to those of the junction between single-band superconductors, see Fig. 2). The dependence $I(\chi)$ is, according to Eq. (7) as follows:

$$I = \frac{\pi|\Delta|}{eR_{N1}}\sin\frac{\chi}{2}\operatorname{sgn}\cos\frac{\chi}{2} - \frac{\pi|\Delta|}{eR_{N2}}\cos\frac{\chi}{2}\operatorname{sgn}\sin\frac{\chi}{2}, \quad (11)$$

and energy of the Josephson junction according to Eq. (8) is equal to

$$E = -\frac{\Phi_0|\Delta|}{2eR_{N1}}\left|\cos\frac{\chi}{2}\right| - \frac{\Phi_0|\Delta|}{2eR_{N2}}\left|\sin\frac{\chi}{2}\right|. \quad (12)$$

Several conclusions follow from Eqs. (11) and (12). Firstl, the Josephson junction becomes frustrated (Fig. 3) with two-fold degenerate ground state

$$X^{(1)} = 2\arctan\left(\frac{R_{N1}}{R_{N2}}\right) \text{ и } X^{(2)} = 2\pi - \arctan\left(\frac{R_{N1}}{R_{N2}}\right). \quad (13)$$

Secondly, the frustrated ground state corresponds to a non-zero phase difference (conventional contact, see. Fig. 2) or π (π-contact). Following the definition of Ref. 35, we call such a Josephson system φ-contact. Thus, the Josephson systems formed by a conventional and a two-band superconductor with the $s_\pm$-wave symmetry of the order parameter leads to a frustrated φ-contact. It should be note that the possibility of frustration of a Josephson tunnel junction between a single-band and a two-band $s_\pm$ superconductors has been predicted earlier within the phenomenological Ginzburg-Landau approach [36].

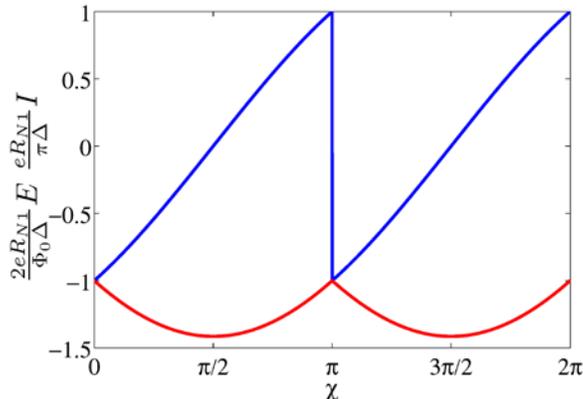

Fig. 3. The current-phase relation (blue line) and the Josephson energy (red line) of a point contact between a single-band and a two-band superconductor with the $s_\pm$-wave symmetry of the order parameter. The ratio $R_{N1}/R_{N2} = 1$.

The behavior of the Josephson junction, which is formed by a single-band and a three-band superconductor, is much more complicated. First of all, this is due to the presence in the three-band superconductor of BTRS phenomenon, which leads to the frustration, namely the emergence of two ground states in the bulk three-band superconductor $\phi_2 = \phi$ and $\phi_3 = \theta$. The values of the phase differences are determined by Eqs. (3)-(6). According to Eq. (7) the Josephson current through the junction is defined as

$$I = \frac{\pi |\Delta|}{eR_{N1}} \sin\frac{\chi}{2} \operatorname{sgn}\cos\frac{\chi}{2} + \frac{\pi |\Delta|}{eR_{N2}} \sin\frac{\chi+\phi}{2} \operatorname{sgn}\cos\frac{\chi+\phi}{2} + \frac{\pi |\Delta|}{eR_{N3}} \sin\frac{\chi+\theta}{2} \operatorname{sgn}\cos\frac{\chi+\theta}{2}, \quad (14)$$

and the Josephson energy is, according to Eq. (8)

$$E = -\frac{\Phi_0 |\Delta|}{2eR_{N1}} \left|\cos\frac{\chi}{2}\right| - \frac{\Phi_0 |\Delta|}{2eR_{N2}} \left|\cos\frac{\chi+\phi}{2}\right| - \left|\cos\frac{\chi+\theta}{2}\right|. \quad (15)$$

To investigate the properties and characteristics of the Josephson junction, the phase differences $\phi$ and $\theta$ can be chosen arbitrarily, since it is always the possible to select such values of the coupling constants $\lambda_{ij}$, that satisfy Eqs. (2)-(6). In other words, after $\phi$ and $\theta$ are selected there are five equations and two inequalities to determine the nine coupling constants. Three equations are self-consistency Eqs. (3)-(6), two others result from the respective equations determining phase differences $\phi$ and $\theta$, two inequalities follow from the second variation of the energy of a bulk three-band superconductor, which determine of the stability of the ground states of the bulk three-band superconductor.

Using these arguments, we consider a three-band superconductor with BTRS by selecting one of ground states in the form $\phi = 0.6\pi$ and $\theta = 1.2\pi$. Since the above phase differences $\phi$ and $\theta$ are in the second and third quadrants, respectively and they belong to the intervals $\phi \in \left[\frac{\pi}{2}, \frac{3\pi}{2}\right]$ and $\theta \in \left[\frac{\pi}{2}, \frac{3\pi}{2}\right]$, and the second ground state corresponds to $\phi = 1.4\pi$ and $\theta = 0.8\pi$.

The dependencies of the current and the Josephson energy on the phase difference $\chi$ of the junction are shown in Fig. 4.

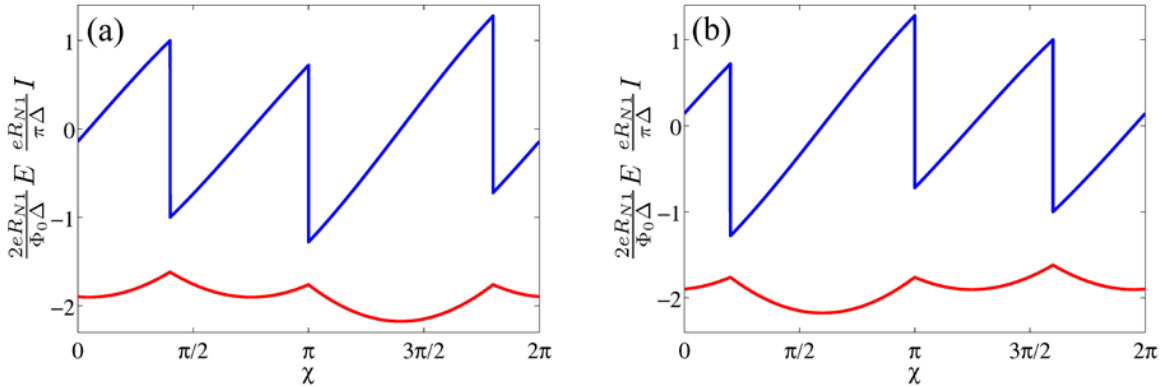

Fig. 4. Current-phase relation (blue line) and the Josephson energy (red line) of a point contact between a single-band and a three-band superconductor with BTRS and frustrated ground state $\phi = 0.6\pi$, $\theta = 1.2\pi$ (**a**) and $\phi = 1.4\pi$, $\theta = 0.8\pi$ (**b**).

As we can see, the frustration of the ground state of a bulk three-band superconductor gives rise to two different dependencies $I(\chi)$ and $E(\chi)$. Practically, this means that in a set of experimental measurements various current-phase relations of the Josephson junction can be observed. Which of them is realized in the specific experiment depends on the history of a three-band superconductor, i.e. in which of the frustrated states is the three-band superconductor during this measurement.

Moreover, the dependence $E(\chi)$ clearly indicates that the Josephson system with a three-band superconductor in the frustrated ground state with $\phi = 0.6\pi$, $\theta = 1.2\pi$ or $\phi = 1.4\pi$, $\theta = 0.8\pi$ behaves like a φ-contact. We also found that such contact with the ballistic conductivity formed by a single-band and a three-band superconductor with BTRS has two local minima on the dependence $E(\chi)$ in addition to the global minimum (Fig. 4).

Now we consider the properties of the Josephson junction, which is formed by a single-band and a three-band superconductors without BTRS, which has the ground states $\phi = 0$, $\theta = \pi$, or $\phi = \pi$, $\theta = \pi$ (case $\phi = 0$, $\theta = 0$ is trivial and is qualitatively matches the properties of a Josephson junction formed by single-band superconductors). Current-phase relations and the energy of the Josephson junction are shown in fig. 5. Although there is no degeneracy of the ground state of a bulk three-band superconductor the Josephson junction undergoes frustration and demonstrates the properties of a φ-contact.

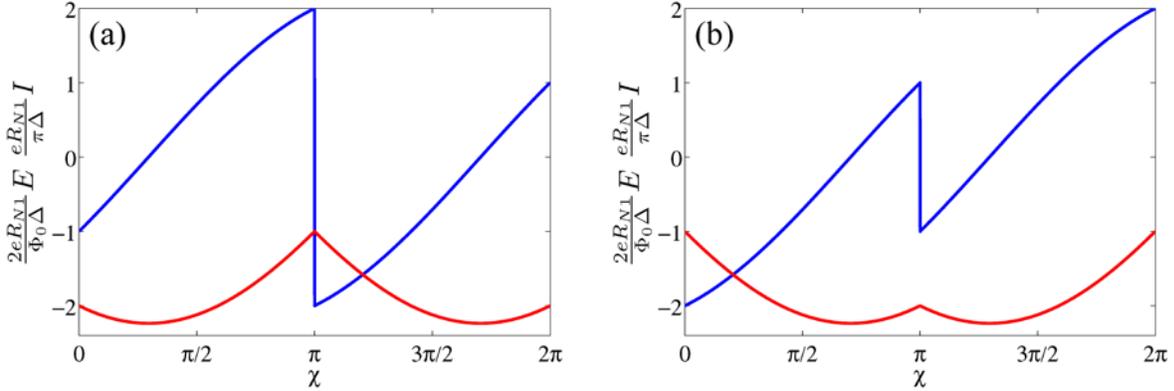

Fig. 5. Current-phase relations (blue line) and the Josephson energy (red line) of a point contact between a single-band and a three-band superconductors without BTRS and ground states $\phi = 0$, $\theta = \pi$ (**a**) and $\phi = \pi$, $\theta = \pi$ (**b**).

Based on Eqs. (14) and (15), the phase diagram of the Josephson junction between a single-band and a three-band superconductor is obtained, which shows the total number of energy minima of the Josephson system depending on the position of the ground state of the bulk three-band superconductor (Fig. 6, left).

As can be seen, the phase diagram is divided into sectors according to the number of energy minima of the Josephson junction. Depending on the values $\phi$ and $\theta$ the number of minima changes from one to three. For each of eleven sectors (I-XI) the position of the minima is given by the expressions in Appendix A.

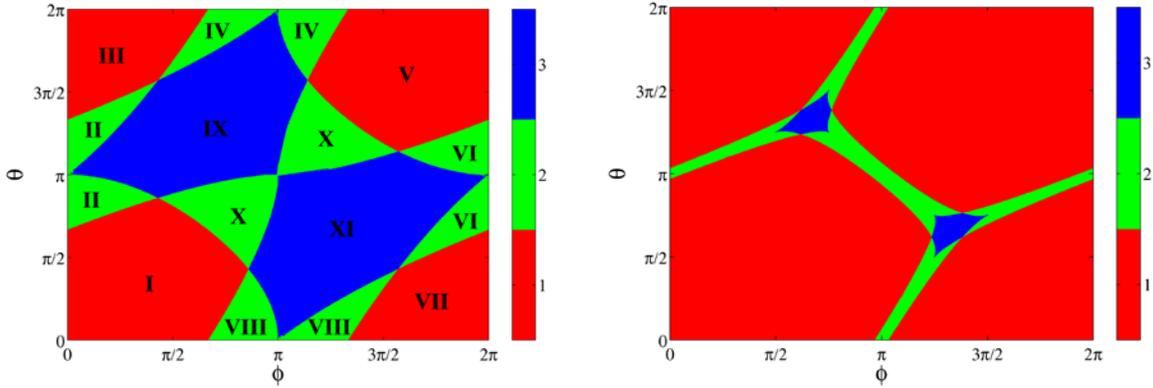

Fig. 6. Dependence of the total number (global and local) of energy minima of a Josephson point contact with ballistic (left) and diffusive (right) conductivity formed by a single-band and a three-band superconductor on the parameters of the ground state ($\phi$ and $\theta$) of the latter.

However, the most notable feature of the Josephson contact with the ballistic conductivity formed by a single-band and a three-band superconductor is much wider variety of the states in comparison with a similar system in the dirty limit (Fig. 6, right). In other words, for the Josephson junction with diffusive conductivity the intervals in which several energy minima exist are significantly narrower (see Appendix B).

## 4. Behaviour of a dc SQUID based on Josephson point contacts between single-band and multi-band superconductors

It is well-known that if one or more of Josephson junctions are included in a superconducting ring, there arise a number of remarkable features related to macroscopic quantum interference phenomenon. In this section we consider these effects in a dc SQUID – a system of two Josephson point contacts that are closed to an s-wave-band and a multi-band superconductor (fig. 7).

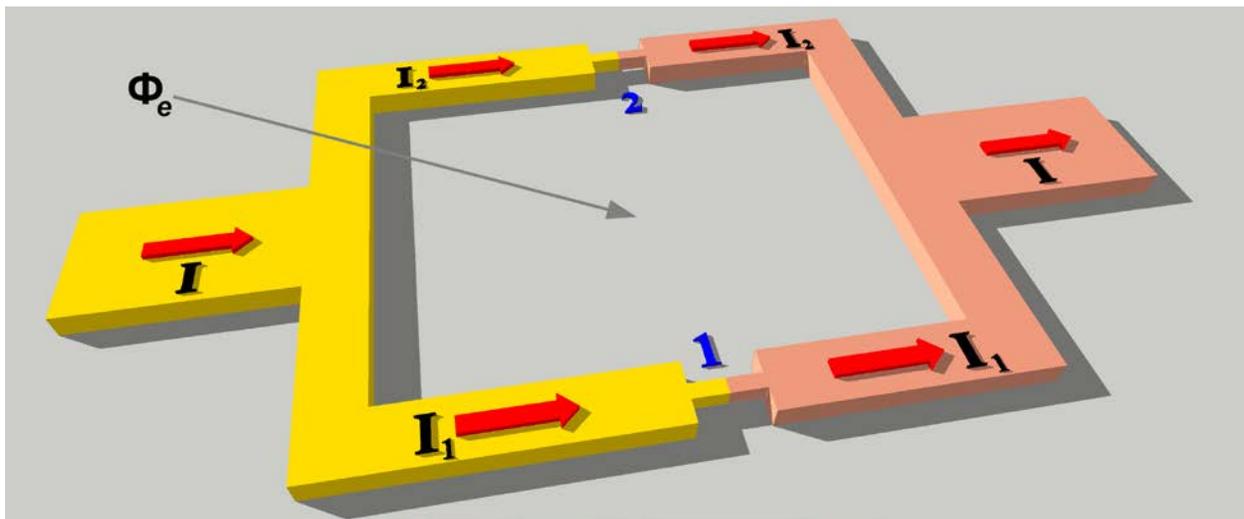

Fig. 7. Schematic model of a dc SQUID based on Josephson point contacts between a single-band (coral color) and multiband (yellow) superconductor with an applied current I and magnetic flux $\Phi_e$. $I_1$ and $I_2$ denote the corresponding currents in point contacts 1 and 2 of the dc SQUID.

Following the above introduced notations, we denote the junction phase difference between the first order parameter of the multi-band superconductor and the order parameter of the single-band superconductor as $\chi_i$, where $i=1,2$ is the contact number. The uniqueness of the phase difference along the contour, the thickness of which is greater than the London penetration depth for both single-band and multi-band superconductors, requires the following the condition to be fulfilled

$$(\chi_1 + \phi_i) - (\chi_2 + \phi_i) = 2\pi \frac{\Phi}{\Phi_0}, \qquad (16)$$

where $\Phi$ is the total magnetic flux through the system and $\Phi_0$ is the magnetic flux quantum. Recall that $\phi_i$ denotes the phase difference between the $i$-th and the first order parameter of the $n$-band superconductor and defines its ground states.

Quantization condition (16) must be supplemented by conditions for the total current and for the total magnetic flux:

$$I = I_1 + I_2, \qquad (17)$$
$$\Phi = \Phi_e + L_1 I_1 - L_2 I_2, \qquad (18)$$

where $I_1$ и $I_2$ are currents flowing through the contacts, $\Phi_e$ is the external magnetic field, $L_1$ and $L_2$ are the inductance of each branch of the dc SQUID. These inductances can be represented as $L_1 = \alpha L$, $L_2 = (1-\alpha)L$, where $L$ is the total dc SQUID ring inductance [37]. Using condition (16), Eqs. (4) and (5) can be rewritten in the dimensionless form:

$$i_1 = (1-\alpha)i + \frac{1}{\beta_{L1}}\left((\chi_1 - \chi_2) - \chi_e\right), \qquad (19)$$

$$i_2 = \alpha i - \frac{1}{\beta_{L1}}\left((\chi_1 - \chi_2) - \chi_e\right), \qquad (20)$$

where the currents $i$, $i_1$ and $i_2$ are expressed now in units of the first-band critical current $I_{c1}^{(1)}$ of the multi-band superconductor without taking into account of the interband interactions for the first point contact (fig. 7), the main parameter of the dc SQUID has the form $\beta_{L1} = \frac{2\pi L I_{c1}^{(1)}}{\Phi_0}$ and the external flux is $\chi_e = \frac{2\pi \Phi_e}{\Phi_0}$.

After introduction of dimensionless variables, the currents $i_1$ and $i_2$ can be expressed as

$$i_j = \sum_i \frac{R_{N1}^{(1)}}{R_{Ni}^{(j)}} \sin \frac{\chi_j + \phi_i \operatorname{sgn}(i-1)}{2} \operatorname{sgn}\left(\cos \frac{\chi_j + \phi_i \operatorname{sgn}(i-1)}{2}\right), \qquad (21)$$

where $R_{Ni}^{(j)}$ are partial contributions of each band of the multi-band superconductor to the normal resistance of the $j$-th contact.

Eqs. (19) and (20) can be obtained from the variation of the energy $E$ with the variables $\chi_j$

$$E(\chi_1, \chi_2) = \frac{1}{2\beta_{L1}}\left((\chi_2 - \chi_1) + \chi_e\right)^2 - i\left[(1-\alpha)\chi_1 + \alpha\chi_2\right] + E_J(\chi_1, \chi_2). \qquad (22)$$

Here $E_J(\chi_1, \chi_2)$ is an expression for the total energy of the Josephson point contacts of the dc SQUID

$$E_J(\chi_1, \chi_2) = -\sum_j \sum_i \frac{R_{N1}^{(1)}}{R_{Ni}^{(j)}} \left|\cos\frac{\chi_j + \phi_i \operatorname{sgn}(i-1)}{2}\right|. \qquad (23)$$

Let us begin the investigation of the dc SQUID behavior by considering its energy as a function of the phase difference $\chi_1$ and $\chi_2$, which depends on the applied magnetic flux. Unless otherwise stated, for the sake of simplicity in the following we will consider symmetric a dc SQUID with identical point contacts $\frac{R_{N1}^{(1)}}{R_{Ni}^{(j)}} = 1$.

Fig. 8 shows the contour plots of the surface $E(\chi_1, \chi_2)$ for dc SQUIDs with Josephson junctions formed between $s$-wave single-band superconductors (Fig. 8a, b), between $s$-wave single-band and $s_\pm$ two-band superconductors (Fig 8c, d), between single-band and BTRS three-band superconductors (Fig.8e, f), between single-band and non-BTRS three-band superconductors (Fig 8g-j) for zero magnetic flux (left column) and the magnetic field corresponding to the half of the flux quantum (right column).

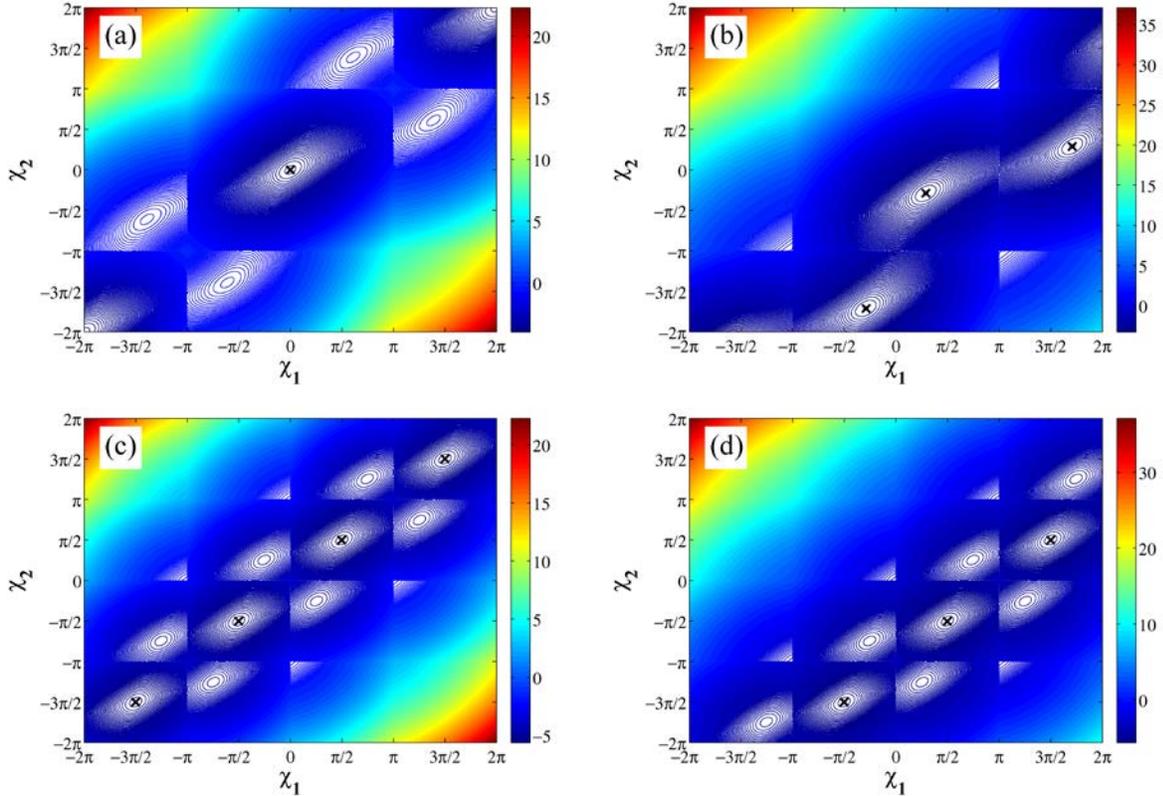

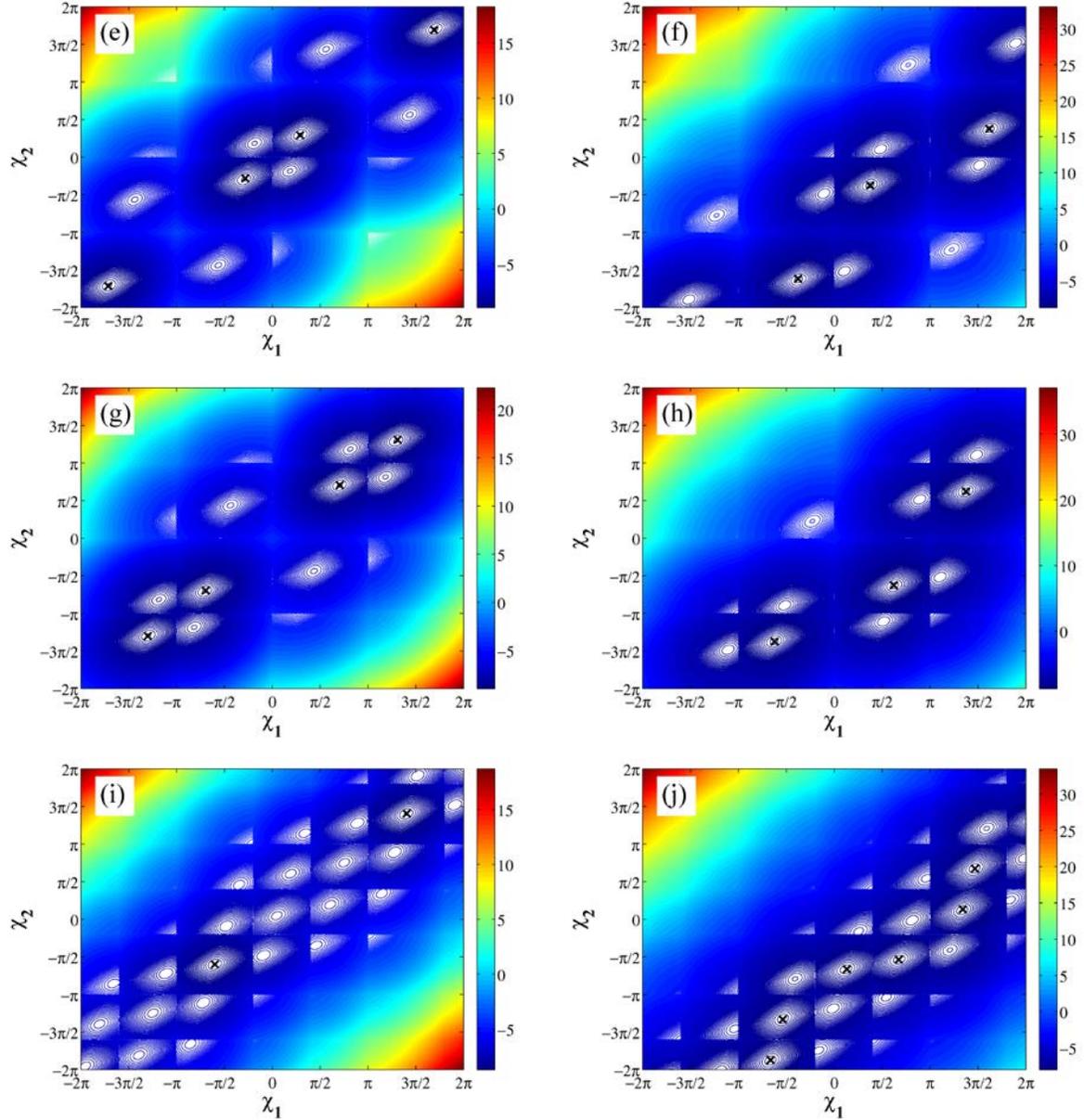

Fig.8. Contour plots of the energy of a dc SQUID for zero external magnetic flux $\chi_e = 0$ ($\Phi/\Phi_0 = 0$, left column) and for the $\chi_e = \pi$ ($\Phi/\Phi_0 = 0.5$, right column) in the absence of bias current. (**a**) and (**b**) are for Josephson point contacts between *s*-wave single-band superconductors; (**c**), (**d**) are for point contacts between *s*-wave single-band and $s_\pm$ two-band superconductors (**e**), (**f**) and (**g**), (**h**) are for point contacts between single-band and non-BTRS three-band superconductors with ground state $\phi = 0, \theta = \pi$ and $\phi = \pi, \theta = \pi$; and (**i**), (**j**) for the frustrated ground state $\phi = 0.6\pi, \theta = 1.2\pi$ of a BTRS three-band superconductor. Crosses indicate positions of the global minima. For all the SQUIDs $\beta_{L1} = 3$.

In the case of a dc SQUID based on a two-band superconductor with $s_\pm$-wave symmetry of the order parameter the energy minimum is degenerate at zero magnetic flux $\Phi_e$ (Fig. 8c) due to the frustration of the Josephson junction (see Fig. 3). For the same reason (see Fig. 5a and b) a degeneracy occurs for a non-BTRS three-band superconductor (Fig 8e and g).

In a magnetic field ($\Phi_e = \Phi_0/2$) dc SQUIDs based on a two-band $s_\pm$ and a non-BTRS three-band superconductor (Fig. 8d, f, h, respectively) demonstrate qualitatively the same behavior as a conventional dc SQUID based on single-band superconductors (Fig. 8b).

The most interesting features emerge for a dc SQUID based on a BTRS three-band superconductor. At zero magnetic flux there is only a shift of the position of the global minimum of energy of the dc SQUID from the zero point $\chi_1 = \chi_2 = 0$ (Fig. 8i), despite the presence of BTRS in the bulk three-band superconductor. However at $\Phi_e = \Phi_0/2$ due to the BTRS state, a unique feature emerges – a strong degeneracy of the energy minimum (Fig. 8j), which is not realized in other dc SQUIDs based on single-band or multi-band superconductors without BTRS (Fig. 8b, d, f, h). Figures 8i and j are contour plots of the energy of a dc SQUID based on a three-band superconductor with the ground state $\phi = 0.6\pi$, $\theta = 1.2\pi$. For a three-band superconductor in another ground state $\phi = 1.4\pi$, $\theta = 0.8\pi$, the above behavior remains qualitatively the same, differing only in a symmetric arrangement of the minima with $\Phi_e = \Phi_0/2$ with respect to the straight line $\chi_1 - \chi_2 = \pi$.

One of the most important characteristics of a SQUID is the dependence of the critical current $i_c$ on the external magnetic flux $\Phi_e$. To simplify the analysis of the problem it is commonly assumed that the inductance of the loop is negligible, so that the total magnetic flux through the SQUID is equal to the external flux

$$\chi_1 - \chi_2 = 2\pi \frac{\Phi_e}{\Phi_0}. \tag{24}$$

In this case the problem of finding the function $i_c = i_c\left(\frac{\Phi_e}{\Phi_0}\right)$ is equivalent to the problem of determining the maximum of the function

$$i(\chi_1, \chi_2) = i_1(\chi_1) + i_2(\chi_2), \tag{25}$$

taking into account the quantization condition (16). Here $i_1(\chi_1)$ and $i_2(\chi_2)$ are the dimensionless current-phase relations defined by Eq. (21).

The dependence of the critical current on the external magnetic flux for a dc SQUID was obtained numerically. This dependence for a dc SQUID, based on Josephson point contacts between single-band superconductors, is shown in Figure 9a. Here and below the critical current is normalized to its maximum value.

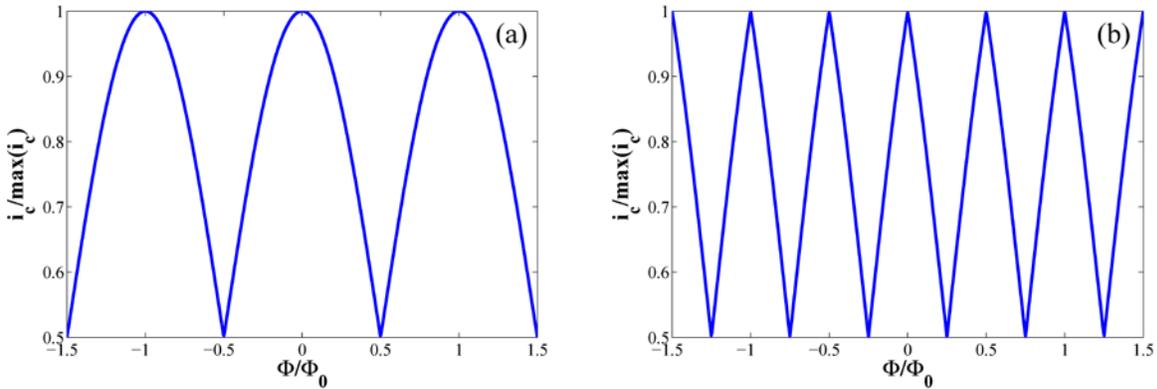

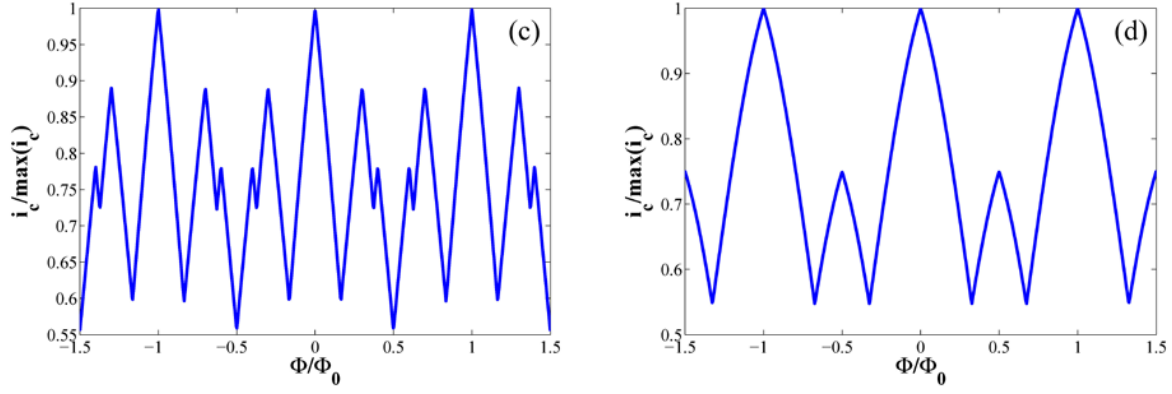

Fig. 9. Dependences of the critical current of a symmetric dc SQUID with vanishing inductance on the applied magnetic flux for a single-band superconductor (**a**), a two-band $s_\pm$ superconductor (**b**), a BTRS three-band superconductor with frustrated ground states $\phi = 0.6\pi, \theta = 1.2\pi$ and $\phi = 1.4\pi, \theta = 0.8\pi$ (**c**) and a non-BTRS three-band superconductor with ground states $\phi = 0, \theta = \pi$ and $\phi = \pi, \theta = \pi$ (**d**).

For a two-band superconductor with s $_{++}$-wave symmetry of the order parameter the dependence $i_c = i_c\left(\dfrac{\Phi_e}{\Phi_0}\right)$ qualitatively agrees with the similar characteristic for a SQUID based on single-band superconductors (see Fig. 9a). Features emerge when the two-band superconductor has $s_\pm$-wave symmetry of the order parameter (Fig. 9b). It can be seen that the critical current has a saw-tooth dependence with the period $\Phi_0/2$ in contrast to the period of $\Phi_0$ for a conventional single-band or a two-band superconductor with $s_{++}$ symmetry.

In the case of a three-band superconductor with BTRS it was found that, despite the presence of two different possible current-phase relations [17], the dependence $i_c = i_c\left(\dfrac{\Phi_e}{\Phi_0}\right)$ is the same for both ground states of a bulk three-band superconductor (Fig. 9c). A similar situation occurs for a three-band superconductor without BTRS, which has the ground state $\phi = 0, \theta = \pi$ and $\phi = \pi, \theta = \pi$ (Fig. 9d).

Comparing figures 9c and 9d we can conclude that the critical current for a three-band superconductor with BTRS has a more complex structure with additional peaks in the dependence $i_c = i_c\left(\dfrac{\Phi_e}{\Phi_0}\right)$.

The incorporation of asymmetry of the critical currents of the Josephson point contacts naturally leads to an asymmetry of dependencies $i_c = i_c\left(\dfrac{\Phi_e}{\Phi_0}\right)$ (fig. 10). This effect is noticeable for a three-band superconductor with BTRS (fig. 10c) and without it (fig. 10d).

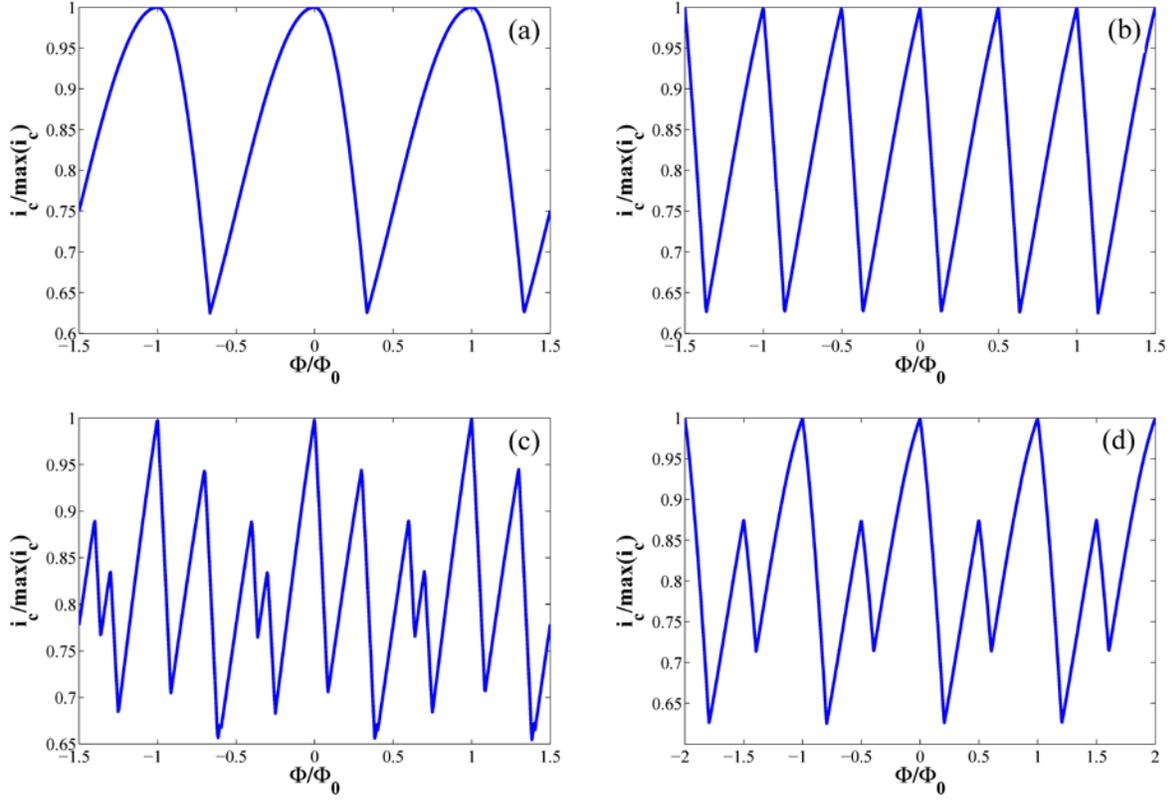

Fig. 10. Dependencies of the critical current of the asymmetrical ($\frac{R_{N1}^{(1)}}{R_{Ni}^{(1)}} = 1$ and $\frac{R_{N1}^{(1)}}{R_{Ni}^{(2)}} = 3$) dc SQUID with vanishing inductance on the applied magnetic flux for a single-band superconductor (**a**), a two-band $s_{\pm}$ superconductor (**b**), a BTRS three-band superconductor with frustrated ground states $\phi = 0.6\pi, \theta = 1.2\pi$ and $\phi = 1.4\pi, \theta = 0.8\pi$ (**c**) and a non-BTRS three-band superconductor with $\phi = 0, \theta = \pi$ and $\phi = \pi, \theta = \pi$ (**d**).

Let us now consider the S-states of a dc SQUID, i.e. the dependencies of the total magnetic flux through the loop on the external magnetic flux for zero bias current $i = 0$. Assuming the critical currents of the Josephson point contacts equal, the dc SQUID can be replaced with the equivalent rf SQUID with a phase difference $\chi_{rf}$ at the contact. This phase difference is related to the phase differences $\chi_1$ and $\chi_2$ of the dc SQUID through the relations

$$\chi_1 = \Delta\chi + \chi_{rf}, \tag{26}$$
$$\chi_2 = \Delta\chi - \chi_{rf}, \tag{27}$$

where $\Delta\chi$ is a certain parameter, which can be found by summing Eqs. (19) and (20) and taking into account the form of the current-phase relation (21)

$$\sum_j \sum_i \sin\frac{\chi_j + \phi_i \operatorname{sgn}(i-1)}{2} \operatorname{sgn}\left(\cos\frac{\chi_j + \phi_i \operatorname{sgn}(i-1)}{2}\right) = 0 \tag{28}$$

and Eqs. (26), (27) for the new variable $\chi_{rf}$.

The value of the parameter $\Delta\chi$ depends on what superconductors are in contact in the dc SQUID. If the dc SQUID is formed by Josephson junctions between single-band superconductors, then in the interval $\Delta\chi \in [0, 2\pi)$ this parameter is

$$\Delta\chi = 0 \text{ and } \Delta\chi = \pi. \tag{29}$$

For a system of Josephson junctions between a single-band and a two-band superconductor with $s_\pm$-wave symmetry of the order parameter

$$\Delta\chi = 0, \; \Delta\chi = \frac{\pi}{2}, \; \Delta\chi = \pi \text{ and } \Delta\chi = \frac{3\pi}{2}. \tag{30}$$

For a three-band superconductor the parameter $\Delta\chi$ depends on $\phi$ and $\theta$ which define the ground states of a bulk three-band superconductor. In this case parameter $\Delta\chi$ can be found only by numerical solution of Eqs. (26) - (28). The exception is the case of a three-band superconductor without BTRS, since its $\Delta\chi$ parameter is the same as for a dc SQUID based on single-band superconductors (29).
Taking into account the change of variables (26) and (27) Eqs. (19) and (20) are transformed into the equation

$$\chi_{rf} + \frac{1}{2}\beta_{L1}\sum_i \sin\frac{\chi_{rf} + \Delta\chi + \phi_i \,\text{sgn}(i-1)}{2} \text{sgn}\left(\cos\frac{\chi_{rf} + \Delta\chi + \phi_i \,\text{sgn}(i-1)}{2}\right) = \frac{1}{2}\chi_e. \tag{31}$$

This equation is transcendental, thus its solutions, namely the functions $\chi_{rf}(\chi_e)$ can be found numerically only. Solving Eq. (31) and taking into account Eqs. (26) and (27) we can find the S-states of a dc SQUID based on the Josephson junctions between two single-band superconductors (fig. 11a), a single-band and a two-band $s_\pm$-wave superconductor (fig. 11b), a single-band and a three-band superconductor with BTRS (fig. 11c) and a single-band and a three-band superconductor without BTRS (Fig. 11d).

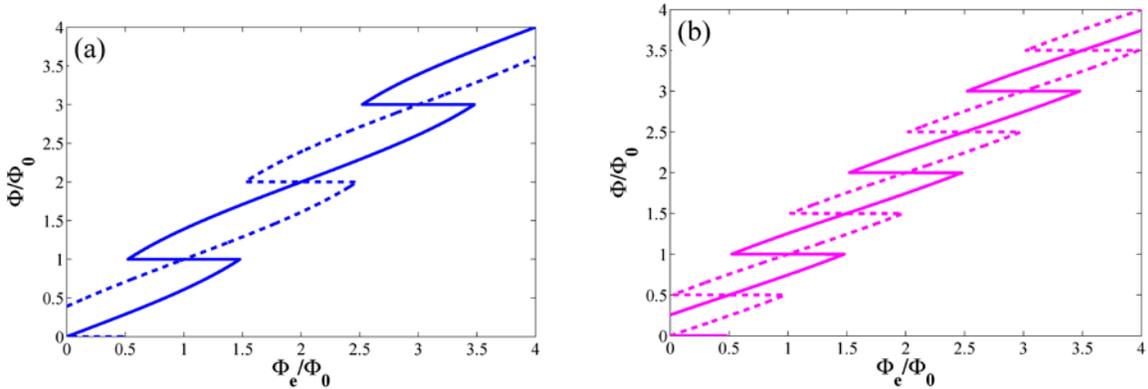

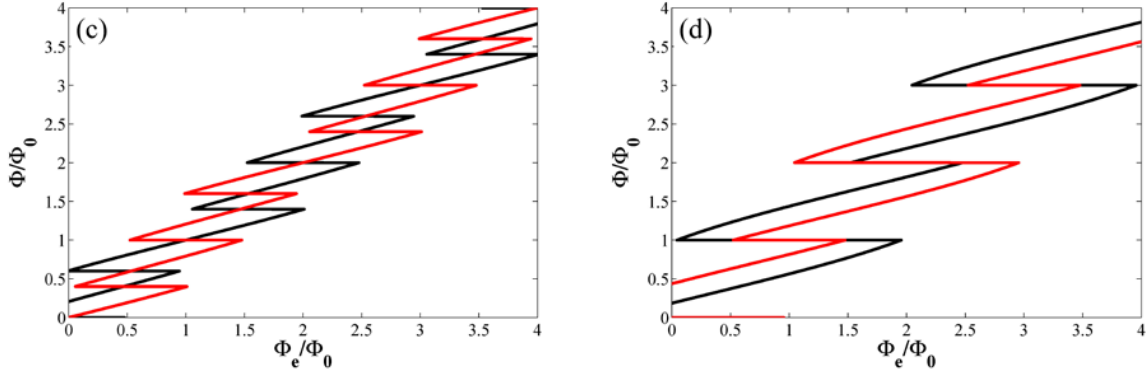

Fig. 11. S states of a conventional dc SQUID (**a**), a dc SQUID based on single-band and two-band $s_\pm$ superconductors (**b**), single-band and BTRS three-band superconductor (**c**), and single-band and non-BTRS three-band superconductors (**d**). S-states for the dc SQUID based on a BTRS three-band superconductor correspond to the ground states $\phi = 0.6\pi, \theta = 1.2\pi$ with $\Delta\chi = 1.25664$ (black line) and $\phi = 1.4\pi, \theta = 0.8\pi$ with $\Delta\chi = \pi - 1.25664$ (red line). In the case of the dc SQUID based on a non-BTRS three-band superconductor S-states correspond to the ground states $\phi = 0, \theta = \pi$ (black line) with $\Delta\chi = 0$ and $\phi = \pi, \theta = \pi$ with $\Delta\chi = 0$ (red line) (**d**). Solid and dash lines in figures (**a**) and (**b**) demonstrate all possible S-states of a conventional dc SQUID and a dc SQUID based on single-band and two-band $s_\pm$ superconductors. The reason for the absent of dotted lines for a dc SQUID based on BTRS and non-BTRS three-band superconductors is given in the main text. For all curves $\beta_{L1} = 3$.

Let us make few comments regarding the S states of a dc SQUID presented in fig. 11. The figure does not show the dependences for a two-band $s_{++}$ superconductor and a three-band superconductor with the ground state $\phi = \theta = 0$, since they qualitatively match the respective characteristics of a dc SQUID based on single-band superconductors (fig. 11a), differing only quantitatively. Despite the fact that Eq. (31) has four solutions for a two-band $s_\pm$-wave superconductor, the S- states for $\Delta\chi = 0$ and $\Delta\chi = \pi$ coincide with each other (solid line in fig. 11b). The same is true for the states $\Delta\chi = \dfrac{\pi}{2}$ and $\Delta\chi = \dfrac{3\pi}{2}$ (dashed line in Fig. 11b).

A similar situation occurs for a dc SQUID based on a BTRS three-band superconductor. Figure 11(c) shows the S states for a dc SQUID based on a three-band superconductor with the frustrated ground state $\phi = 0.6\pi$, $\theta = 1.2\pi$ and $\phi = 1.4\pi$, $\theta = 0.8\pi$ are shown on fig. 11c. Obviously, in this case four different $\Phi(\Phi_e)$ dependences are possible, however for the ground state $\phi = 0.6\pi$, $\theta = 1.2\pi$, the S-states with the parameters $\Delta\chi = 1.25664$ and $\Delta\chi = 1.25664 + \pi$ coincide pairwise with the S-states for $\phi = 1.4\pi$, $\theta = 0.8\pi$ with the parameters $\Delta\chi = 2\pi - 1.25664$ and $\Delta\chi = \pi - 1.25664$. The same pattern can be implemented for other three-band superconductors with BTRS.

For a three-band superconductor without BTRS the S-states for $\phi = 0$, $\theta = \pi$ with the parameters $\Delta\chi = 0$ and $\Delta\chi = \pi$ ( (Fig. 11d) are identical to the S-states for $\phi = \pi$, $\theta = \pi$ with the parameters $\Delta\chi = \pi$ and $\Delta\chi = 0$, respectively.

As in the case of a conventional dc SQUID, the S states are stable only if the derivative $d\Phi/d\Phi_e > 0$. In other words, the S states are stable in the intervals where dependences $\Phi(\Phi_e)$ have a positive slope.

In comparison with the hysteretic behavior of a dc SQUID based on single-band superconductors (Fig. 11a) the S states of dc SQUIDs based on multi-band superconductors can demonstrate a multi-hysteretic behavior. Moreover, the greater is the number of energy gaps, the more jumps can be detected when measuring the $\Phi(\Phi_e)$ dependence. As Fig. 11(c) implies, the largest number of hysteresis loops can be observed in a three-band superconductor with BTRS.

Therefore, the presence of the S states in a dc SQUID can be regarded as a kind of necessary condition for the possible existence of multi-band superconductivity and the realization of BTRS phenomena in such multi-band superconductors.

## Conclusions

In summary, we have studied the properties of Josephson systems based on point contacts with ballistic conductivity formed between a single-band and a multiband (two- and three-band) superconductor at zero temperature. The ballistic Josephson point contact between a single-band and an $s_\pm$-wave superconductor is frustrated, has two ground states, and thus demonstrates the properties of a φ-contact. If the Josephson junction is formed by a single-band and a three-band superconductor with BTRS, such a contact may have two different current-phase relations, also demonstrating the φ-contact properties. Furthermore, depending on the ground state of the three-band superconductor with BTRS, the Josephson junction can have from one to three energy minima. These minima can be either all stable in the global sense (three-fold degeneracy of the ground state) or only one of them can be globally stable. For a three-band superconductor, which is characterized by the absence of BTRS, the Josephson junction has qualitatively the same properties as the contact with an $s_\pm$-wave two-band superconductors, it is a frustrated (two-fold degenerate) φ-contact. It was found that in comparison with the Josephson junction with a diffusive conductivity, the ballistic junction between a single-band and a three-band superconductor can demonstrate a significantly wider variety of states with additional local or global energy minima.

We also considered the behavior of a dc SQUID based on the Josephson junctions formed by a single-band and a multi-band superconductor. We found the differences in the characteristics of dc SQUIDs (the dependences of the critical current and the S-state) constructed of an s±-wave superconductor, a three-band superconductor with BTRS and a three-band superconductor without BTRS as compared with conventional dc SQUID based on single-band superconductors. The above features can be used to detect the presence of a multi-band structure in superconductors. Moreover, in the case of a three-band superconductor these results can help to detect BTRS.

This work was supported by a grant from the DKNII (M/231-2013) and a grant from the BMBF (UKR-2012-028). One of the authors (Y.Y.) was supported by a grant from the RSF No. 15-12-10020.

## Appendix A

The phase diagram (fig. 6a) consists of 11 sectors, each of which corresponds to the total number of energy minima of the Josephson ballistic junction formed by a single-band and a three-band superconductor. The position of each minimum (local and/or global) determined by following equations:

Sector I (one minimum)

$$X^{(1)} = 2\pi - 2\operatorname{arccot} \frac{1 + 2\cos\frac{\phi+\theta}{4}\cos\frac{\phi-\theta}{4}}{2\sin\frac{\phi+\theta}{4}\cos\frac{\phi-\theta}{4}}, \tag{A1}$$

Sector II (two minima)

$$X^{(1)} = 2\pi - 2\text{arccot}\frac{1 + 2\cos\frac{\phi+\theta}{4}\cos\frac{\phi-\theta}{4}}{2\sin\frac{\phi+\theta}{4}\cos\frac{\phi-\theta}{4}}, \tag{A2}$$

$$X^{(2)} = 2\text{arccot}\frac{1 + 2\sin\frac{\theta+\phi}{4}\sin\frac{\theta-\phi}{4}}{2\sin\frac{\theta-\phi}{4}\cos\frac{\theta+\phi}{4}}, \tag{A3}$$

Sector III (one minimum)

$$X^{(1)} = \begin{cases} 2\text{arccot}\dfrac{1 + 2\sin\frac{\theta+\phi}{4}\sin\frac{\theta-\phi}{4}}{2\sin\frac{\theta-\phi}{4}\cos\frac{\theta+\phi}{4}}, & \text{if } \phi+\theta < 2\pi \\[2ex] 2\pi + 2\text{arccot}\dfrac{1 + 2\sin\frac{\theta+\phi}{4}\sin\frac{\theta-\phi}{4}}{2\sin\frac{\theta-\phi}{4}\cos\frac{\theta+\phi}{4}}, & \text{if } \phi+\theta > 2\pi, \end{cases} \tag{A4}$$

Sector IV (two minima)

$$X^{(1)} = 2\pi + 2\text{arccot}\frac{1 - 2\cos\frac{\phi+\theta}{4}\cos\frac{\phi-\theta}{4}}{2\sin\frac{\phi+\theta}{4}\cos\frac{\phi-\theta}{4}}, \tag{A5}$$

$$X^{(2)} = 2\text{arccot}\frac{1 + 2\sin\frac{\theta+\phi}{4}\sin\frac{\theta-\phi}{4}}{2\sin\frac{\theta-\phi}{4}\cos\frac{\theta+\phi}{4}}, \tag{A6}$$

Sector V (one minimum)

$$X^{(1)} = 2\text{arccot}\frac{1 - 2\cos\frac{\phi+\theta}{4}\cos\frac{\phi-\theta}{4}}{2\sin\frac{\phi+\theta}{4}\cos\frac{\phi-\theta}{4}}, \tag{A7}$$

Sector VI (two minima)

$$X^{(1)} = 2\text{arccot}\frac{1 - 2\cos\frac{\phi+\theta}{4}\cos\frac{\phi-\theta}{4}}{2\sin\frac{\phi+\theta}{4}\cos\frac{\phi-\theta}{4}}, \tag{A8}$$

$$X^{(2)} = 2\pi + 2\operatorname{arccot} \frac{-1 + 2\sin\frac{\theta+\phi}{4}\sin\frac{\theta-\phi}{4}}{2\sin\frac{\theta-\phi}{4}\cos\frac{\theta+\phi}{4}}, \tag{A9}$$

Sector VII (one minimum)

$$X^{(1)} = \begin{cases} 2\operatorname{arccot} \dfrac{-1 + 2\sin\frac{\theta+\phi}{4}\sin\frac{\theta-\phi}{4}}{2\sin\frac{\theta-\phi}{4}\cos\frac{\theta+\phi}{4}}, & \text{if } \phi+\theta < 2\pi \\[2ex] 2\pi + 2\operatorname{arccot} \dfrac{-1 + 2\sin\frac{\theta+\phi}{4}\sin\frac{\theta-\phi}{4}}{2\sin\frac{\theta-\phi}{4}\cos\frac{\theta+\phi}{4}}, & \text{if } \phi+\theta > 2\pi \end{cases} \tag{A10}$$

Sector VIII (two minima)

$$X^{(1)} = 2\pi - 2\operatorname{arccot} \frac{1 + 2\cos\frac{\phi+\theta}{4}\cos\frac{\phi-\theta}{4}}{2\sin\frac{\phi+\theta}{4}\cos\frac{\phi-\theta}{4}}, \tag{A11}$$

$$X^{(2)} = 2\operatorname{arccot} \frac{-1 + 2\sin\frac{\theta+\phi}{4}\sin\frac{\theta-\phi}{4}}{2\sin\frac{\theta-\phi}{4}\cos\frac{\theta+\phi}{4}}, \tag{A12}$$

Sector IX (three minima)

$$X^{(1)} = 2\pi - 2\operatorname{arccot} \frac{1 + 2\cos\frac{\phi+\theta}{4}\cos\frac{\phi-\theta}{4}}{2\sin\frac{\phi+\theta}{4}\cos\frac{\phi-\theta}{4}}, \tag{A13}$$

$$X^{(2)} = 2\operatorname{arccot} \frac{1 - 2\cos\frac{\phi+\theta}{4}\cos\frac{\phi-\theta}{4}}{2\sin\frac{\phi+\theta}{4}\cos\frac{\phi-\theta}{4}}, \tag{A14}$$

$$X^{(3)} = \begin{cases} 2\operatorname{arccot} \dfrac{1 + 2\sin\frac{\theta+\phi}{4}\sin\frac{\theta-\phi}{4}}{2\sin\frac{\theta-\phi}{4}\cos\frac{\theta+\phi}{4}}, & \text{if } \phi+\theta < 2\pi \\[2ex] 2\pi + 2\operatorname{arccot} \dfrac{1 + 2\sin\frac{\theta+\phi}{4}\sin\frac{\theta-\phi}{4}}{2\sin\frac{\theta-\phi}{4}\cos\frac{\theta+\phi}{4}}, & \text{if } \phi+\theta > 2\pi, \end{cases} \tag{A15}$$

Sector X (two minima)

$$X^{(1)} = 2\pi - 2\text{arccot}\frac{1+2\cos\frac{\phi+\theta}{4}\cos\frac{\phi-\theta}{4}}{2\sin\frac{\phi+\theta}{4}\cos\frac{\phi-\theta}{4}}, \tag{A16}$$

$$X^{(2)} = 2\text{arccot}\frac{1-2\cos\frac{\phi+\theta}{4}\cos\frac{\phi-\theta}{4}}{2\sin\frac{\phi+\theta}{4}\cos\frac{\phi-\theta}{4}}, \tag{A17}$$

Sector XI (three minima)

$$X^{(1)} = 2\pi - 2\text{arccot}\frac{1+2\cos\frac{\phi+\theta}{4}\cos\frac{\phi-\theta}{4}}{2\sin\frac{\phi+\theta}{4}\cos\frac{\phi-\theta}{4}}, \tag{A18}$$

$$X^{(2)} = 2\text{arccot}\frac{1-2\cos\frac{\phi+\theta}{4}\cos\frac{\phi-\theta}{4}}{2\sin\frac{\phi+\theta}{4}\cos\frac{\phi-\theta}{4}}, \tag{A19}$$

$$X^{(3)} = \begin{cases} 2\text{arccot}\dfrac{-1+2\sin\frac{\theta+\phi}{4}\sin\frac{\theta-\phi}{4}}{2\sin\frac{\theta-\phi}{4}\cos\frac{\theta+\phi}{4}}, & \text{if } \phi+\theta < 2\pi \\[2ex] 2\pi + 2\text{arccot}\dfrac{-1+2\sin\frac{\theta+\phi}{4}\sin\frac{\theta-\phi}{4}}{2\sin\frac{\theta-\phi}{4}\cos\frac{\theta+\phi}{4}}, & \text{if } \phi+\theta > 2\pi \end{cases} \tag{A20}$$

# Appendix B

The current-phase relation $I(\chi)$ and the energy of the Josephson junction $E(\chi)$ formed by a single-band and a three-band superconductor in the dirty limit is given by [32]

$$\begin{aligned} I = & \frac{\pi|\Delta|}{eR_{N1}}\cos\frac{\chi}{2}\text{arctanh}\sin\frac{\chi}{2} + \frac{\pi|\Delta|}{eR_{N2}}\cos\frac{\chi+\phi}{2}\text{arctanh}\sin\frac{\chi+\phi}{2} \\ & + \frac{\pi|\Delta|}{eR_{N3}}\cos\frac{\chi+\theta}{2}\text{arctanh}\sin\frac{\chi+\theta}{2}, \end{aligned} \tag{B1}$$

$$E = \frac{|\Delta|\Phi_0}{2eR_{N1}}\left(2\sin\frac{\chi}{2}\operatorname{arctanh}\sin\frac{\chi}{2} + \ln\cos^2\frac{\chi}{2}\right) + \frac{|\Delta|\Phi_0}{2eR_{N2}}\left(2\sin\frac{\chi+\phi}{2}\operatorname{arctanh}\sin\frac{\chi+\phi}{2} + \ln\cos^2\frac{\chi+\phi}{2}\right)$$
$$+ \frac{|\Delta|\Phi_0}{2eR_{N3}}\left(2\sin\frac{\chi+\theta}{2}\operatorname{arctanh}\sin\frac{\chi+\theta}{2} + \ln\cos^2\frac{\chi+\theta}{2}\right),$$
(B2)

where $\phi$ and $\theta$ define the ground state of a bulk three-band superconductor.

The phase diagram in Figure 6 (right), which demonstrates the number of energy minima of the Josephson diffusive junction as a function of $\phi$ and $\theta$, is based on the Eq. (B1) and (B2).